\documentclass[aps,prb,english,footinbib,floatfix,reprint,twocolumn,showpacs,superscriptaddress,longbibliography,10pt,tightenlines]{revtex4-1}
\usepackage[utf8]{inputenc}
\usepackage{amssymb,amsmath}
\usepackage{graphicx}
\usepackage{natbib}
\usepackage{multirow}
\usepackage{braket}
\usepackage{subcaption}
\usepackage{ragged2e}
\DeclareCaptionJustification{justified}{\justifying}
\usepackage[final]{hyperref} 
\hypersetup{
	colorlinks=true,       
	linkcolor=red,        
	citecolor=red,        
	filecolor=magenta,     
	urlcolor=blue         
}



\begin{document}

\title{Relaxation dynamics in the (double) sine-Gordon model - an open-system viewpoint}

\author{D. Szász-Schagrin}\email{szaszdavid@gmail.com}
\affiliation{Department of Theoretical Physics, Institute of
Physics, Budapest University of Technology and
Economics, M{\H u}egyetem rkp. 3., H-1111 Budapest,
Hungary}
\affiliation{BME-MTA Statistical Field Theory ’Lend\"ulet’ Research
Group, Budapest University of Technology and
Economics, M{\H u}egyetem rkp. 3., H-1111 Budapest,
Hungary}
\author{D. X. Horváth}\email{esoxluciuslinne@gmail.com}
\affiliation{SISSA and INFN Sezione di Trieste, via Bonomea 265, 34136 Trieste, Italy}
\affiliation{Department of Mathematics, King’s College London, Strand, London WC2R 2LS, United Kingdom}
\author{G. Takács}\email{takacs.gabor@ttk.bme.hu}
\affiliation{Department of Theoretical Physics, Institute of
Physics, Budapest University of Technology and
Economics, M{\H u}egyetem rkp. 3., H-1111 Budapest,
Hungary}
\affiliation{BME-MTA Statistical Field Theory ’Lend\"ulet’ Research
Group, Budapest University of Technology and
Economics, M{\H u}egyetem rkp. 3., H-1111 Budapest,
Hungary}
\affiliation{MTA-BME Quantum Dynamics and Correlations
Research Group, Budapest University of Technology and
Economics, M{\H u}egyetem rkp. 3., H-1111 Budapest,
Hungary}
\date{August 23, 2024}

\begin{abstract}
    We study the effects of integrability breaking on the relaxation dynamics of the (double) sine-Gordon model. Compared to previous studies, we apply an alternative viewpoint motivated by open-system physics by separating the phase field into homogeneous and inhomogeneous parts, describing a quantum pendulum (subsystem) and an interacting phononic bath (environment). To study the relaxation dynamics in the model, we perform quantum quenches using the mini-superspace-based truncated Hamiltonian approach developed recently and simulate the real-time evolution of various entanglement measures and the energy transfer between the subsystem and its environment. Our findings demonstrate that in the presence of integrability-breaking perturbations, the relaxation dynamics is substantially faster, signalled by the increase of entanglement and energy transfer between the quantum pendulum and the phonon bath.
\end{abstract}

\maketitle

\section{Introduction}

Out-of-equilibrium dynamics of isolated quantum many-body systems has been at the forefront of research in recent years \cite{2010NJPh...12e5006C,2011RvMP...83..863P,2016RPPh...79e6001G,2016JSMTE..06.4001C}. Engineering and manipulating isolated quantum systems has now become routine in experiments with trapped ultracold atoms \cite{2006Natur.440..900K,2007Natur.449..324H,2012NatPh...8..325T,2012Sci...337.1318G,2013NatPh...9..640L,2013PhRvL.111e3003M,2013Natur.502...76F,2015Sci...348..207L}, which opened the way to experimental studies of the relaxation and thermalisation of closed quantum systems. A paradigmatic protocol is the quantum quench\cite{2006PhRvL..96m6801C,2007JSMTE..06....8C}, which corresponds to a sudden change of some parameters of a system which is initially prepared in an equilibrium state. 

The sine-Gordon model considered in this work is a fundamental example of interacting integrable quantum field theories. It also provides an effective description of the low-energy physics of numerous physical systems, such as, e.g., spin chains\cite{1999PhRvB..60.1038A, 2009PhRvB..79r4401U, 2004PhRvL..93b7201Z}, circuit quantum eletrodynamics\cite{2019PhRvB.100o5425R, 2021NuPhB.96815445R}, and bosonic and fermionic Hubbard models\cite{2005odhm.book.....E, Giamarchi:743140, Nagerl2010, 2000cond.mat.11439C}. Due to its integrability, many equilibrium properties of the model are exactly known, ranging from exact results on scattering amplitudes and form factors to expectation values of local observables\cite{Zamolodchikov:1978xm,1992ASMP...14.....S,1997NuPhB.493..571L}. It has also attracted interest in the context of non-equilibrium dynamics due to experimental realisation with two Josephson-coupled one-dimensional bosonic quasi-condensates \cite{2013NatPh...9..640L,2017Natur.545..323S}. In the experiment, ultra-cold atoms are trapped in an elongated double-well potential, limiting the physics to one spatial dimension. The effective description of the system can be obtained using bosonisation\cite{2007PhRvB..75q4511G}, predicting that the anti-symmetric modes of the double-well potential realise the sine-Gordon model, weakly coupled to a Luttinger liquid accounting for symmetric modes, which lead to breaking of integrability. In general, integrability-breaking couplings are always expected to be present in any experimental realisation, and this observation serves as the main motivation of the present study.

In this paper, avoiding the introduction of additional degrees of freedom (which are inevitably present in experiments), we model integrability breaking within the space of the sine-Gordon scalar field. Since the definition of the field as a (relative) bosonic quasi-condensate phase variable dictates the periodicity of the potential, the only relevant operators which also preserve field parity are higher cosine harmonics. Therefore, we model integrability-breaking effects by adding a cosine term with a period twice the fundamental one. 

A natural interpretation of the sine-Gordon system is that of a quantum pendulum defined by the field zero mode nonlinearly coupled to an interacting gas of phonons made up by the non-zero modes. We consider the dynamics in this picture and investigate mode-mode energy transfer, especially between the zero-mode pendulum and the interacting phonon gas, and their mutual entanglement, commonly used when studying the relaxation in open quantum systems. This viewpoint is reminiscent of open dissipative quantum systems, the paradigmatic example of which is the Caldeira-Leggett (CL) model \cite{1983PhyA..121..587C}. In the case of the sine-Gordon model, splitting the system into the quantum pendulum and the phononic bath is close in spirit, if not in all details, to the original CL setting. We investigate this system using the recently developed minisuperspace-based truncated Hamiltonian approximation (MSTHA) \cite{2024PhRvB.109a4308S}. The truncated Hamiltonian approach to quantum field theory started with the seminal work of Yurov and Zamolodchikov \cite{Yurov_1989,1991IJMPA...6.4557Y} who developed it to study relevant perturbations of conformal field theories. It was later extended to various other models \cite{Feverati_1998,2014JSMTE..12..010C,Rychkov_2015,2015NuPhB.899..547K,Bajnok_2016,Konik2021JHEP}, and also applied to the out-of-equilibrium dynamics of various field theories \cite{Rakovszky_2016, Horv_th_2017,Horvath2019,Horvath2022in, Horvath2022}. The variant applied here, called MSTHA, is specially adapted to the open-system picture described before. It also allows for the investigation of the effect of integrability breaking on relaxation in the sine-Gordon model. Recently, the effect of integrability breaking on dissipative open quantum systems dynamics has attracted interest \cite{2019PhRvL.123y4101A}. Furthermore, the above approach is also natural in connection with the recently developed tensor network-based Hamiltonian truncation \cite{2023arXiv231212506S}.

The outline of the paper is as follows. In Section \ref{sec:system} and \ref{sec:protocol_observables}, we introduce the system and the quench protocol, respectively. Section \ref{sec:dynamics} presents our results for the dynamics, with zero-mode entanglement considered in Subsection \ref{subsec:entanglement} and energy transfer in Subsection \ref{subsec:energy_transfer}. We summarise our results and present our conclusions in \ref{sec:conclusion}. 

\section{The sine-Gordon model as an open system}\label{sec:system}

Time evolution in the sine-Gordon model is governed by the Hamiltonian
\begin{equation}
    \hat{H}_\textrm{sG} = \int dx:\left(\frac{1}{2}(\partial_t\hat{\varphi})^2 + \frac{1}{2}(\partial_x\hat{\varphi})^2 - \lambda\cos\beta\hat{\varphi} \right):\,.
     \label{eq:sg-ham}
\end{equation}
Here, the semi-colons denote normal ordering with respect to modes of the free massless boson corresponding to $\lambda = 0$. Throughout this paper, we set the parameter $\beta$ (characterising the strength of the interactions) to $0<\beta<\sqrt{4\pi}$, called the attractive regime of the sine-Gordon model. For these values of $\beta$, apart from the solitonic excitations, the theory admits quasi-particles in the form of stable bound states of solitons called breathers that have discrete spectrum:
\begin{equation}
    m_n = 2 M\sin\frac{\pi\xi n}{2}, \quad \xi = \frac{\beta^2}{8\pi-\beta^2}\,,
\end{equation}
where $M$ is the soliton mass and $n$ is an integer less than $1/\xi$. Integrability allows for determining the exact relation of the coupling $\lambda$ to the mass scale, which we choose to be the lightest breather mass $m_1$\cite{1995IJMPA..10.1125Z}:
\begin{equation}
    \lambda = \kappa m_1^{2-2\Delta}
\end{equation}
with    
\begin{equation}
\kappa = \left(2\sin\frac{\pi\xi}{2}\right)^{2\Delta-2}\frac{2\Gamma(\Delta)}{\pi\Gamma(1-\Delta)}\left(\frac{\sqrt{\pi}\Gamma\left(\frac{1}{2-2\Delta}\right)}{2\Gamma\left(\frac{\Delta}{2-2\Delta}\right)}\right)^{2-2\Delta}\label{eq:lambda}
\end{equation}
where $\Delta$ is related to the anomalous dimension of the cosine operator:
\begin{equation}
    2\Delta = \frac{\beta^2}{4\pi}\,.
\end{equation}
The MSTHA requires formulating the theory in a finite volume $L$. Recalling that the field $\hat\varphi$ is an angular variable with a period of $2\pi/\beta$, a natural choice of boundary conditions is the following quasi-periodic form:
\begin{equation}
    \hat\varphi(x + L, t) = \hat\varphi(x, t) + \frac{2\pi}{\beta}m \label{eq:pbc}
\end{equation}
where $m$ is an integer called the winding number (or topological charge). In our subsequent calculations, we only consider the sector corresponding to $m = 0$ for which Eq.~\eqref{eq:pbc} reduces to the usual periodic boundary conditions.

Usually, the Hamiltonian \eqref{eq:sg-ham} is considered a relevant perturbation of the compactified free massless boson put in a finite volume $L$:
\begin{equation}
    \hat{H}_\text{FB} = \int_0^L dx:\left(\frac{1}{2}(\partial_t\hat{\varphi})^2+\frac{1}{2}(\partial_x\hat{\varphi})^2\right):\,.
    \label{eq:cftham}
\end{equation}

Here, we adopt a different point of view introduced in a previous work \cite{2024PhRvB.109a4308S}. The field $\hat\varphi$ can be separated into the homogeneous part (the zero-wave number Fourier mode) and the inhomogeneous part (non-zero modes)
\begin{equation}
    \hat{\varphi}(x,t) = \hat{\varphi}_0(t) + \tilde{{\varphi}}(x,t)\,.\label{eq:field-decomp}
\end{equation}
Neglecting the spatial dependence of $\hat\varphi$ (that is, the contribution of the non-zero modes) gives the single-mode description of the theory, and the sine-Gordon Hamiltonian \eqref{eq:sg-ham} reduces to that of a quantum pendulum \cite{2024PhRvB.109a4308S}:
\begin{equation}
    \hat{H}_\text{QP} = \frac{1}{2L}\hat{\pi}_0^2 -\lambda L \left(\frac{2\pi}{L }\right)^{2\Delta}\cos(\beta\hat{\varphi}_0)\,,
    \label{eq:qp-ham}
\end{equation}
where $\hat\pi_0$ is the zero mode of the canonical conjugate momentum to $\hat\varphi$ satisfying the bosonic computational relation $[\hat{\varphi}_0, \hat{\pi}_0] = i$. The full sine-Gordon model describes a quantum pendulum coupled to a set of non-linear, interacting \textit{phononic mode}s (the oscillator modes):
\begin{align}
    \hat{H}_{\text{sG}} =\frac{1}{2}\int_0^L:\left[(\partial_t\hat{\varphi}_0)^2 + (\partial_t{\tilde{\varphi}})^2+(\partial_x{\tilde{\varphi}})^2\right]: \nonumber\\- \frac{\lambda}{2}\int_0^L dx:\left[e^{i\beta\hat{\varphi}_0}e^{i\beta\tilde{\varphi}} + e^{-i\beta\hat{\varphi}_0}e^{-i\beta\tilde{\varphi}}\right]:\,.\label{eq:sg-ham-decomposed}
\end{align}
From this expression alone, the physical meaning of the phononic degrees of freedom is ambiguous, allowing for some flexibility in their definitions. The inhomogeneous part $\tilde\varphi(x,t)$ in Eq.~\eqref{eq:field-decomp} is a sum of the non-zero modes of the field:
\begin{equation}
    \tilde\varphi(x,t) = \sum_{k\neq 0}\hat\varphi_k(t) e^{-ik\frac{2\pi}{L}x}\,.
\end{equation}
In our definition, each mode $\hat\varphi_{k}$ describes a harmonic oscillator 
\begin{equation}
    \hat H_k = L\left[\frac{1}{2}\left(\dot{\hat{\varphi}}_{k}\dot{\hat{\varphi}}_{-k}+\dot{\hat{\varphi}}_{-k}\dot{\hat{\varphi}}_{k}\right) + \frac{1}{2}\omega_k^2 \left({\hat{\varphi}}_{k}{\hat{\varphi}}_{-k}+{\hat{\varphi}}_{-k}{\hat{\varphi}}_{k}\right)\right]
\end{equation}
with frequency
\begin{equation}
    \omega_k^2 = \frac{4\pi^2 k^2}{L^2} + \mu^2 
\end{equation}
with
\begin{equation}
    \mu^2 = \kappa\left(\frac{2\pi}{l}\right)^{2\Delta}\beta^2\,,
\end{equation}
which simply follows from the Taylor expansion of the cosine potential up to second order \footnote{For the volume dependence, see the work \cite{2024PhRvB.109a4308S}.}, while the parameter $l = m_1 L$ is the dimensionless volume. 

In this approach, the sine-Gordon model describes a quantum pendulum coupled to a bath of interacting harmonic oscillators (phonons):
\begin{equation}
    \hat H_{\rm sG} = \hat H_{\rm QP} + \sum_{k\neq 0}\hat H_k + \hat H_{\rm int}\,,\label{eq:open-qp}
\end{equation}
where the second term corresponds to the Hamiltonian of the oscillator bath, and the interaction term $H_{\rm int}$ is a complicated expression describing both the interaction between the pendulum and the bath and the self-interactions of the bath itself.

At first glance, this redefinition of the various degrees of freedom might seem superfluous. However, this open-system viewpoint enables us to study the relaxation of the model in terms of the subsystem (the quantum pendulum) and the environment (the phononic bath) and apply concepts such as the energy transfer between the subsystem and the environment or the dynamics of their mutual entanglement. Moreover, since the sine-Gordon model describes the full system, its dynamics is integrable. A naturally arising question is the effect of integrability breaking on the relaxation dynamics of the subsystem.

To study the integrability breaking of the model, we perturb the sine-Gordon Hamiltonian \eqref{eq:sg-ham} by an additional cosine term, yielding the double sine-Gordon model
\begin{align}
      \hat{H}_\textrm{dsG} = \int dx:\bigg(\frac{1}{2}(\partial_t\hat{\varphi})^2 + \frac{1}{2}(\partial_x\hat{\varphi})^2 &- \lambda\cos\beta\hat{\varphi}\nonumber \\ 
      &-\bar\alpha\cos2\beta\hat{\varphi}\bigg):\,.
       \label{eq:dsg-ham}
\end{align}
The strength of integrability breaking is characterised by the dimensionless parameter
\begin{equation}
       \alpha = \bar{\alpha}m_1^{2\Delta_2 - 2};\quad\quad \Delta_2 = 4\Delta\,.
       \label{eq:alpha_def}
\end{equation}
Breaking the integrability of the sine-Gordon theory can be achieved in several ways; however, preserving the $2\pi/\beta$ periodicity of the field and the $\phi\rightarrow-\phi$ symmetry limits the perturbations to higher harmonics of the cosine potential. As a result, a natural choice for the integrability-breaking term is a double-frequency cosine, which is expected to be the leading one in generic situations.
Additionally, to preserve the vacuum structure of the theory, we limit ourselves to suitably small values of $\alpha$; more precisely, to $\alpha$ values for which the classical potential does not develop new local minima:
\begin{equation}
    \bar\alpha < \bar\alpha^*;\quad \partial_\varphi^2\left(-\lambda \cos\beta\varphi - \bar\alpha^*\cos 2\beta\varphi\right)|_{\varphi = \pi/\beta} = 0\,,
\end{equation}
that is
\begin{equation}
    \bar\alpha < \frac{\lambda}{4}\,.
\end{equation}

The open-system viewpoint adopted previously for the sine-Gordon model \eqref{eq:open-qp} naturally carries over to its non-integrable counterpart:
\begin{equation}
    \hat H_{\rm dsG} = \hat H_{\rm QP} + \sum_{k\neq 0}\hat H_k + \hat H_{\rm int}\,,\label{eq:open-qp-dsg}
\end{equation}
with two main differences. Firstly, the interaction term $\hat H_{\rm int}$ now also contains integrability-breaking perturbations. Secondly, the phonon Hamiltonians $\hat H_{k}$ now describe harmonic oscillators with frequency
\begin{equation}
    \omega_k^2 = \frac{4\pi^2 k^2}{L^2}  + \mu^2 
\end{equation}
with
\begin{equation}
    \mu^2 = \kappa\beta^2\left(\frac{2\pi}{l}\right)^{2\Delta} + 4\alpha\beta^2\left(\frac{2\pi}{l}\right)^{2\Delta_2} \,,
\end{equation}
and the additional last term is the contribution from the double-frequency cosine, with $\alpha$ defined in Eq.~\eqref{eq:alpha_def}.

\section{Quench protocol and observables}\label{sec:protocol_observables}

\subsection{Quench protocol: initial states}\label{subsec:initial_states}
The effect of integrability breaking in the theory can be studied by simulating quantum quenches and examining the relaxation dynamics. After a quantum quench, the time evolution of some initial state $\ket{\Psi_0}$ is governed by the (double) sine-Gordon model:
\begin{equation}
    \ket{\Psi(t)} = e^{-iH_{\rm dsG}t}\ket{\Psi_0}\,.
\end{equation}
The magnitude of the quantum quench can be characterised by the induced energy density 
\begin{equation}
    \frac{1}{L}\left(\braket{\Psi_0|\hat H_{\rm dsG}|\Psi_0} - \braket{\Omega_\alpha|\hat H_{\rm dsG}|\Omega_\alpha}\right)\label{eq:inj-en}
\end{equation}
during the quench, previously used to demonstrate the effectiveness of the MSTHA \cite{2024PhRvB.109a4308S}. Here, $\ket{\Omega_\alpha}$ is the ground state of the double sine-Gordon model for a fixed $\alpha$. A quench is considered weak (or small) if the injected energy density is much smaller than the mass squared of the lightest post-quench quasi-particle $m_1^2(\alpha)$; in the opposite case, it is considered strong. We study a representative initial state for both weak and strong quenches:

\begin{itemize}
    \item For the weak quenches (small injected energy density): the initial state denoted as $\ket{\Psi_{\rm QP}}$ corresponds to preparing both the quantum pendulum and the phononic modes in their respective ground states.
    \item For the strong quenches (large injected energy density): the initial state denoted as $\ket{\Psi_{\rm FB}}$ is chosen to be the ground state of the massless free boson \eqref{eq:cftham}. In the open-system viewpoint, this corresponds to the phononic oscillators prepared in their respective ground state, while the quantum pendulum is in a highly excited state. 
\end{itemize}

The second quench protocol has the additional advantage of being experimentally realisable in cold-atomic experiments, corresponding to a suddenly turning on tunnelling between two initially decoupled one-dimensional bosonic quasi-condensates\cite{2018PhRvA..97d3611H,Horvath2019}. Moreover, this scenario was previously studied in detail using various numerical methods\cite{2024PhRvB.109a4308S,Horvath2019} with an emphasis on determining the validity of semi-classical approaches.

\subsection{Tracking the relaxation dynamics}\label{subsec:observables}

Here, we describe the quantities we use to track the relaxation dynamics. The first class of quantities characterises the mutual entanglement between the quantum pendulum and the phonon bath: the entanglement entropy and the linear entropy. Both can be computed from the reduced density matrix of the quantum pendulum, which is defined as
\begin{equation}
    \hat\rho_0(t) = \rm{Tr}_{\{k\}}\ket{\Psi(t)}\bra{\Psi(t)}\,,
\end{equation}
where $\rm{Tr}_{\{k\}}$ denotes the partial trace over all non-zero modes, which can be straightforwardly computed within the MSTHA from the time-evolving state $\ket{\Psi(t)}$.

\begin{figure}[h]
    \begin{center}
    \begin{subfigure}[b]{0.5\textwidth}
        \centering
        \includegraphics{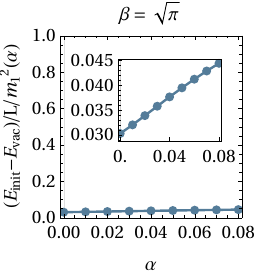}\includegraphics{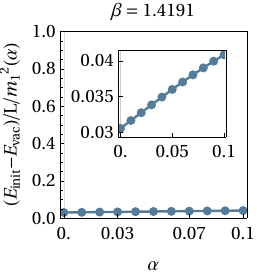}
         \caption{Quantum pendulum ground state}
    \end{subfigure}
    \begin{subfigure}[b]{0.5\textwidth}
        \centering
        \includegraphics{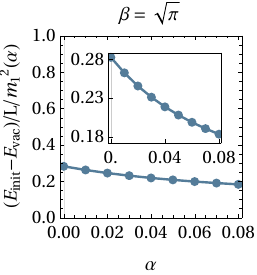}\includegraphics{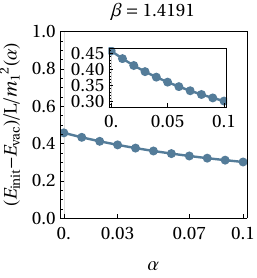}
        \caption{Free boson ground state}
    \end{subfigure}
    \caption{The injected energy density during the quench relative to the mass squared of the post-quench first breather mass $m_1^2(\alpha)$ displayed as a function of the integrability breaking parameter $\alpha$. The natural scale for this quantity is the interval $[0, 1]$, shown in the main figures. The same is displayed on the insets, with the $y$-axis scale chosen so that the features of the curves are more visible. Whenever the energy density is much smaller than $m_1^2(\alpha)$, the quench is considered weak.}
    \label{fig:en-inj}
    \end{center}
\end{figure}

\begin{table}[h]
    \centering
    \begin{tabular}{|c|c|c|}
    \hline
         & $\ket{\Psi_{\rm QP}}$ & $\ket{\Psi_{\rm FB}}$ \\
         \hline
       $\beta = \sqrt{\pi}$  & 32247 & 616923\\
       \hline
       $\beta = 1.4191$  & 25081 &388433\\
    \hline
    \end{tabular}
    \caption{The truncated Hilbert space dimensions used for the different quench scenarios (columns) and coupling parameters (rows).}
    \label{tab:dims}
\end{table}

The entanglement entropy between the subsystem and its environment is given by
\begin{equation}\label{eq:entanglement-entropy}
    S_0(t) = -\rm{Tr}\,\hat\rho_0(t)\log\hat\rho_0(t)\,,
\end{equation}
which is the standard measure of entanglement for pure states and quantifies the system-bath correlations in a way independent of any particular observable.

Another useful measure is given by the purity $\mathcal{P}$\cite{PhysRevD.107.106007}:
\begin{equation}
    \mathcal{P} = \rm{Tr}\,\hat\rho_0^2(t)\,,
\end{equation}
which is also directly measurable in experiments \cite{2015Natur.528...77I}. In our subsequent calculations, we focus on the opposite of the purity, called the linear entropy $S_l$, which characterises the degree of mixedness of the state of the quantum pendulum:
\begin{equation}
    S_l = 1 - \mathcal{P}\,.\label{eq:lin-entropy}
\end{equation}

Apart from various entanglement measures, a key quantity in the relaxation dynamics is the rate of the energy transfer between the subsystem and its environment, which we quantify by computing the real-time mode-mode energy transfer following quantum quenches. We can compute the total energy injected into the system by the quench, as well as the time-dependent energies stored in the zero-mode quantum pendulum and the phononic modes by subtracting the post-quench vacuum energy from the expectation values of the relevant Hamiltonian:
\begin{align}
    &\mathcal{E}_{\Psi}^\textrm{tot} = \braket{\Psi|H_\text{dsG}|\Psi} - \braket{\Omega_\alpha|H_\text{dsG}|\Omega_\alpha}\nonumber\\
    &\mathcal{E}_{\Psi}^{(0)} = \braket{\Psi|H_\text{QP}|\Psi} - \braket{\Omega_\alpha|H_\text{QP}|\Omega_\alpha}\\
    &\mathcal{E}_{\Psi}^{(k)} = \braket{\Psi|H_k|\Psi} - \braket{\Omega_\alpha|H_k|\Omega_\alpha}\nonumber\,.\label{eq:shifted-energies}
\end{align}
The time evolution can then be characterised by the ratios of energy stored in the quantum pendulum and the phononic modes:
\begin{equation}
\begin{split}
    &\varepsilon_{\Psi}^{(0)} = \frac{\mathcal{E}_{\Psi}^{(0)}}{\mathcal{E}_{\Psi}^\textrm{tot}} = \frac{\braket{\Psi|H_\text{QP}|\Psi} - \braket{\Omega_\alpha|H_\text{QP}|\Omega_\alpha}}{\braket{\Psi|H_\text{dsG}|\Psi} - \braket{\Omega_\alpha|H_\text{dsG}|\Omega_\alpha}}\\
    &\varepsilon_{\Psi}^{(k)} = \frac{\mathcal{E}_{\Psi}^{(k)}}{\mathcal{E}_{\Psi}^\textrm{tot}} = \frac{\braket{\Psi|H_k|\Psi} - \braket{\Omega_\alpha|H_k|\Omega_\alpha}}{\braket{\Psi|H_\text{dsG}|\Psi} - \braket{\Omega_\alpha|H_\text{dsG}|\Omega_\alpha}}\,.
\end{split}\label{eq:normed-shifted-energies}
\end{equation}
The left-over portion
\begin{equation}
    \varepsilon_{\Psi}^{\rm int}=1-\varepsilon_{\Psi}^{(0)} - \sum_k \varepsilon_{\Psi}^{(k)}
    \label{eq:int-energy}
\end{equation}
corresponds to the energy stored in the interaction term of the Hamiltonian as written in Eq.~\eqref{eq:open-qp-dsg}.

\section{Dynamics following dsG quenches}\label{sec:dynamics}

\begin{figure*}[h!]
    \begin{center}
    \begin{subfigure}[b]{\textwidth}
        \centering
        \includegraphics{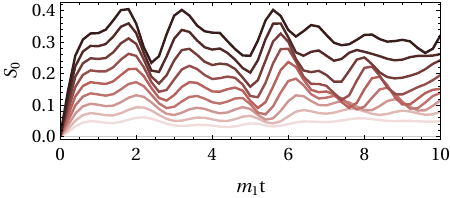}\includegraphics{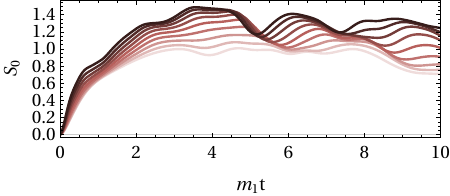}\includegraphics{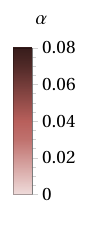}
         \caption{$\beta = \sqrt{\pi}$.}
        \label{fig:entanglement-entropy-qp}
    \end{subfigure}
    \begin{subfigure}[b]{\textwidth}
        \centering
        \includegraphics{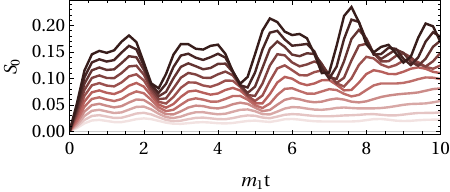}\includegraphics{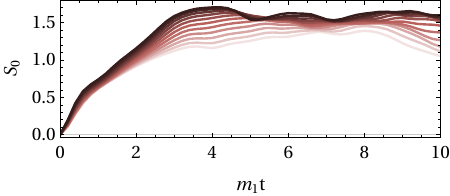}\includegraphics{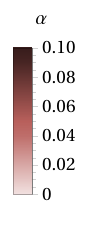}
        \caption{$\beta = 1.4191$}
        \label{fig:entanglement-entropy-fb}
    \end{subfigure}
    \caption{The entanglement entropy \eqref{eq:entanglement-entropy} dynamics following quantum quenches from the two different initial states: $\ket{\Psi_{\rm QP}}$ (left) and $\ket{\Psi_{\rm FB}}$ (right) for various values of the integrability breaking parameter $\alpha$ and interaction strengths $\beta$.}
    \label{fig:entanglement-entropy}
    \end{center}
\end{figure*}

We now turn to our results for the entanglement dynamics and energy transfer following quantum quenches. To simulate the time evolution, we apply the recently developed MSTHA \cite{2024PhRvB.109a4308S}, particularly suited to this problem. The method can be well controlled by analysing the convergence of the results by changing the truncation parameter. All presented results show minimal or no truncation effects and can be considered accurate up to $\sim 1\%$. The dimensions of truncated Hilbert space used in our calculations are shown in Table \ref{tab:dims}. As mentioned, we work in units $m_1 = 1$ and a finite volume, which we fix as $l = m_1L = 10$. As a result, we consider relatively short time scales $m_1 t\lesssim 10.$, where the finite size effects can be neglected. Moreover, since the relaxation times are generally expected to grow as $\beta$ is decreased, in the following, we set $\beta = \sqrt{\pi} \approx 1.7725$ and $\beta = 1.4191$ that correspond to relatively strong interactions. As it turns out, these time scales are sufficient to observe the effects of integrability breaking in all the studied quantities.

\begin{figure*}[h!]
    \begin{center}
    \begin{subfigure}[b]{\textwidth}
        \centering
        \includegraphics{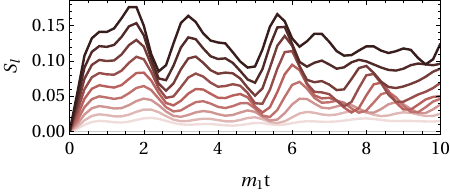}\includegraphics{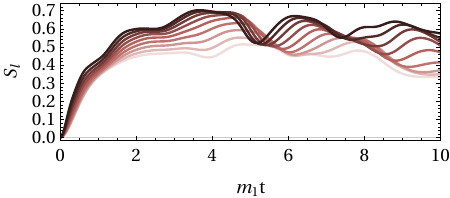}\includegraphics{leg008.pdf}
         \caption{$\beta = \sqrt{\pi}$}
         \label{fig:linear-entropy-qp}
    \end{subfigure}
    \begin{subfigure}[b]{\textwidth}
        \centering
        \includegraphics{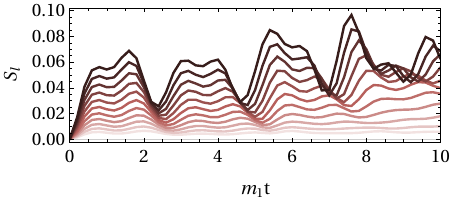}\includegraphics{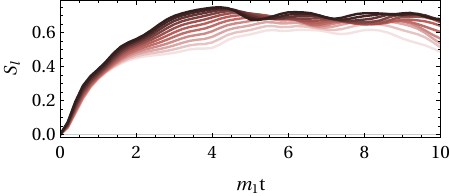}\includegraphics{leg01.pdf}
        \caption{$\beta = 1.4191$}
        \label{fig:linear-entropy-fb}
    \end{subfigure}
    \caption{The time evolution of the linear entropy \eqref{eq:lin-entropy} following quantum quenches from the two different initial states: $\ket{\Psi_{\rm QP}}$ (left) and $\ket{\Psi_{\rm FB}}$ (right) for various values of the integrability breaking parameter $\alpha$ and interaction strengths $\beta$.}
    \label{fig:linear-entropy}
    \end{center}
\end{figure*}

\begin{figure*}[t]
    \begin{center}
    \begin{subfigure}[b]{\textwidth}
        \centering
        \includegraphics{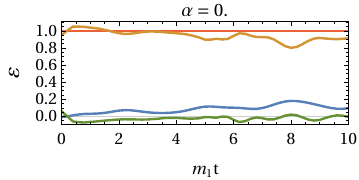}\includegraphics{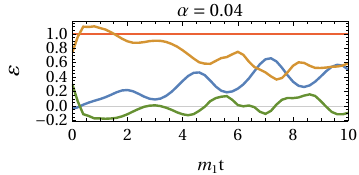}\includegraphics{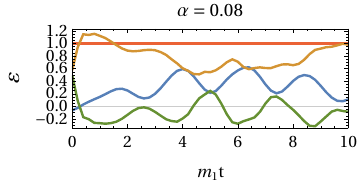}
         \caption{Initial state is the quantum pendulum ground state.}
         \label{fig:energy-transfer1-qp}
    \end{subfigure}
    \begin{subfigure}[b]{\textwidth}
        \centering
       \includegraphics{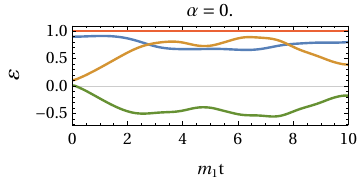}\includegraphics{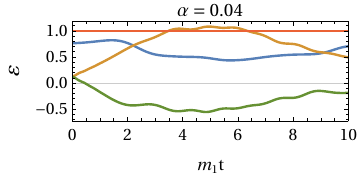}\includegraphics{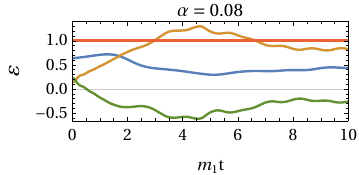}
        \caption{Initial state is the free boson ground state.}
        \label{fig:energy-transfer1-fb}
    \end{subfigure}
    \caption{The mode-mode energy transfer following quantum quenches for $\beta = \sqrt{\pi}$ and various values of the integrability breaking parameter $\alpha$. The lines show the energy content \eqref{eq:normed-shifted-energies} of the quantum pendulum (blue), the phonon gas (orange) and the interaction term \eqref{eq:int-energy} (green) normalised by the total energy (shown in red, by definition equal to 1).}
    \label{fig:energy-transfer1}
    \end{center}
\end{figure*}

\begin{figure*}[t]
    \begin{center}
    \begin{subfigure}[b]{\textwidth}
        \centering
        \includegraphics{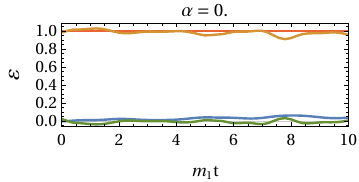}\includegraphics{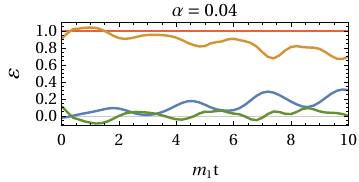}\includegraphics{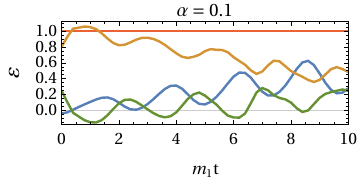}
         \caption{Initial state is the quantum pendulum ground state.}
         \label{fig:energy-transfer2-qp}
    \end{subfigure}
    \begin{subfigure}[b]{\textwidth}
        \centering
       \includegraphics{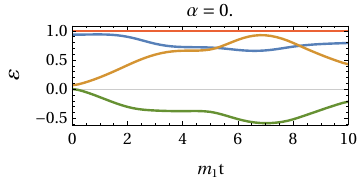}\includegraphics{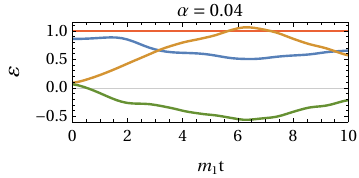}\includegraphics{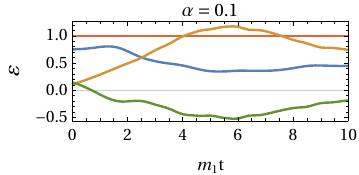}
        \caption{Initial state is the free boson ground state.}
        \label{fig:energy-transfer2-fb}
    \end{subfigure}
    \caption{The mode-mode energy transfer following quantum quenches for $\beta = 1.4191$ and various values of the integrability breaking parameter $\alpha$. The lines show the energy content \eqref{eq:normed-shifted-energies} of the quantum pendulum (blue), the phonon gas (orange) and the interaction term \eqref{eq:int-energy} (green) normalised by the total energy (shown in red, by definition equal to 1).}
    \label{fig:energy-transfer2}
    \end{center}
\end{figure*}

In Fig. \ref{fig:en-inj}, the injected energy density is shown relative to the mass squared of the post-quench first breather $m_1^2(\alpha)$. As mentioned before, the quench scenario when the system is initiated in the free boson ground state corresponds to a much larger energy density than that of the quantum pendulum ground state with a difference of an order of magnitude. The latter is considered a very weak quench, with limited excess energy available to excite particles. Moreover, as the integrability breaking parameter $\alpha$ increases, the injected energy density increases only minimally for the case of the quantum pendulum ground state. At the same time, it even decreases in the case of the free boson ground state. As a result, the observed changes in the relaxation (presented later in this section) can not be attributed to the larger energy densities present and are a clear effect of the integrability breaking.

\subsection{Entanglement measures}\label{subsec:entanglement}

In Fig. \ref{fig:entanglement-entropy}, the entanglement entropy \eqref{eq:entanglement-entropy} dynamics is shown following our two types of quantum quenches for $\beta = \sqrt{\pi}$ and $1.4191$ and multiple values of the integrability breaking parameter $\alpha$. In both scenarios, the subsystem is initially in a pure state, and the entanglement entropy is zero. For short times, there is a fast saturation to some maximal entropy value with accompanying oscillations. For the stronger quench shown in Fig. \ref{fig:entanglement-entropy-fb}, the saturation value is much larger compared to the weak quench showcased in Fig. \ref{fig:entanglement-entropy-qp}, signalling stronger entanglement between the subsystem and the bath.

The effect of integrability breaking on the system's relaxation is also very clear. Irrespective of the quench setup, increasing the integrability breaking parameter $\alpha$ results in stronger oscillations and faster saturation of the entanglement entropy. The saturation values also grow with $\alpha$ and correspond to a stronger entanglement between the subsystem and its environment. 

This picture remains largely unchanged when considering the time evolution of the linear entropy of the quantum pendulum, displayed in Fig. \ref{fig:linear-entropy}. At $t = 0$, the subsystem is prepared in a pure state and the linear entropy is zero. For finite times, the linear entropy behaves very similarly to the entanglement entropy and quickly saturates to some value that is much larger for the stronger quench. The saturation value and the relaxation speed grow substantially with increasing the strength of integrability breaking $\alpha$, corresponding to the increase in the mixedness of the subsystem due to integrability breaking.

\subsection{Energy transfer}\label{subsec:energy_transfer}

Now we turn to discussing the results concerning the time evolution of the energy transfer. To characterise the relaxation of the system, we focus on the ratios of energy stored in the quantum pendulum, the phononic modes and the interaction term \eqref{eq:normed-shifted-energies}-\eqref{eq:int-energy}. 

In Figs. \ref{fig:energy-transfer1} and \ref{fig:energy-transfer2} the energy transfer can be seen following quantum quenches for $\beta = \sqrt{\pi}$ and $\beta = 1.4149$, respectively. The data are shown for three different values of $\alpha$, corresponding to the ends and the middle of the studied interval. \footnote{Data for the remaining $\alpha$ parameters are available in animated form from the preprint at \href{https://arxiv.org/abs/2408.14428}{arXiv.org}.}

For the weak quench scenario, where the injected energy density during the quench is small, the mode-mode energy transfer is generally very slow, even for the relatively strong couplings studied here. In contrast, the energy transfer is much faster in the strong quench scenario with higher injected energy density, as apparent in Figs. \eqref{fig:energy-transfer1-fb} and \eqref{fig:energy-transfer2-fb}. In this case, the increase in the energy transfer is accompanied by a suppression of oscillations as well.

The energy transfer is the slowest for $\alpha = 0$ when the theory is integrable — gradually increasing the strength of integrability breaking $\alpha$ results in the increase of the energy transfer, irrespective of the quench scenario studied. Additionally, the amplitude of the fluctuations also grows when the injected energy density is small. Apart from these oscillations, the energy contained in the interaction terms is limited: the majority of the energy is contained in the bath, except for short times in the case of free boson ground state, where the originally highly excited quantum pendulum carries most of the energy content of the system.

These findings demonstrate that breaking the integrability of the theory results in a substantial increase in the energy transfer between the subsystem and its environment, which translates to an increased relaxation rate in the theory.

\section{Conclusion}\label{sec:conclusion}

Relaxation of isolated quantum many-body systems is a problem that has been very actively studied in contemporary research. While generic non-integrable systems are expected to thermalise, equilibration in integrable systems is much more complicated. Therefore, a key question is the effects of the breakdown of integrability on the relaxation dynamics in these systems.

To study the effects of integrability breaking on the relaxation dynamics, we performed quantum quenches in the double sine-Gordon model, a non-integrable deformation of the paradigmatic sine-Gordon model, which is a relativistic integrable quantum field theory formulated in terms of a periodic scalar field $\hat\varphi$. Here, we considered this problem from a less common open-system viewpoint by decomposing the field into homogeneous and inhomogeneous parts, which describe a quantum pendulum and a bath of interacting harmonic oscillators, respectively. This alternative viewpoint allowed us to study relaxation as the dynamics of a subsystem (the quantum pendulum) coupled to an environment (the oscillator bath), which made it natural to study various entanglement measures and the mode-resolved energy transfer.

We simulated two classes of quantum quenches, characterised by the magnitude of their injected energy densities. The simulation used the recently developed mini-superspace-based truncated Hamiltonian approach (MSTHA) \cite{2024PhRvB.109a4308S}, which is especially well-suited for this problem and allows us to obtain very accurate results for the time evolution.

For both quench scenarios, we find a clear effect of the integrability breaking on the relaxation dynamics: increasing the strength of integrability breaking, the relaxation rate also increases, as apparent from the faster energy transfer and growing entanglement between the quantum pendulum and its environment. Apart from the total entanglement and mixedness of the subsystem, the rate at which the entanglement and linear entropies grow also increases, quickly saturating to their respective maximum values. These findings are irrespective of the quench scenarios studied; however, for small quenches, the initially very slow energy transfer is much more affected by the integrability breaking compared to the strong quench.

Apart from the relevance of our results to existing cold-atomic experiments \cite{2013NatPh...9..640L,2017Natur.545..323S}, the open-system viewpoint also provides interesting future directions, such as generalisation to other systems to study the relaxation dynamics or phenomena such as parametric resonances in the theory\cite{2022PhRvB.106g5426L}.

\begin{acknowledgments}
D.X.H. is grateful to S. Sotiriadis for inspiring and insightful discussions. This work was supported by the National Research, Development and Innovation Office of Hungary (NKFIH) through OTKA Grant No. ANN 142584. The NKFIH also provided partial support to D.S.-S. by the scholarship Grant No. NKP-23-3-II-BME-182, and to G.T. via the Grant “Quantum Information National Laboratory“ with Grant No. 2022-2.1.1-NL-2022-00004. D.X.H. acknowledges support also from ERC under Consolidator Grant No. 771536 NEMO and from the Engineering and Physical Sciences Research Council (EPSRC) under Grant No. EP/W010194/1. Finally, D.X.H. is grateful for accessing the resources of KCL’s CREATE HPC cluster.
\end{acknowledgments}

\bibliographystyle{utphys}
\bibliography{dsg_quench}
\end{document}